\documentclass[aps,pra,twocolumn,showpacs,superscriptaddress,floatfix]{revtex4}
\usepackage{graphicx}
\usepackage{epsfig}

\usepackage{longtable}
\usepackage{amssymb}
\usepackage{amsthm}
\usepackage{amsfonts}
\usepackage[centertags]{amsmath}
\usepackage{color}
\usepackage{sidecap}
\usepackage{pst-all}
\usepackage{pstricks}
\usepackage[latin1]{inputenc}
\usepackage[english]{babel}
\usepackage{soul}
\usepackage{pifont} 
\pdfoutput=1

\usepackage{color} 

\newcommand{\beq}{\begin{equation}}
\newcommand{\eeq}{\end{equation}}
\newcommand{\bea}{\begin{eqnarray}}
\newcommand{\eea}{\end{eqnarray}}

\newcommand{\ket}[1]{\left|\,#1\,\right\rangle}  
\newcommand{\vari}[1]{\Delta^2{#1}}

\newcommand{\coon}{(Color online) }

\begin{document}


\title{Experimental characterization of frequency dependent squeezed light}

\author{Simon Chelkowski}
\affiliation{Institut f\"ur Atom- und Molek\"ulphysik,
Universit\"at Hannover and Max-Planck-Institut f\"ur
Gravitationsphysik (Albert-Einstein-Institut), Callinstr. 38,
30167 Hannover, Germany}

\author{Henning Vahlbruch}
\affiliation{Institut f\"ur Atom- und Molek\"ulphysik,
Universit\"at Hannover and Max-Planck-Institut f\"ur
Gravitationsphysik (Albert-Einstein-Institut), Callinstr. 38,
30167 Hannover, Germany}

\author{Boris Hage}
\affiliation{Institut f\"ur Atom- und Molek\"ulphysik,
Universit\"at Hannover and Max-Planck-Institut f\"ur
Gravitationsphysik (Albert-Einstein-Institut), Callinstr. 38,
30167 Hannover, Germany}

\author{Alexander Franzen}
\affiliation{Institut f\"ur Atom- und Molek\"ulphysik,
Universit\"at Hannover and Max-Planck-Institut f\"ur
Gravitationsphysik (Albert-Einstein-Institut), Callinstr. 38,
30167 Hannover, Germany}

\author{Nico Lastzka}
\affiliation{Institut f\"ur Atom- und Molek\"ulphysik,
Universit\"at Hannover and Max-Planck-Institut f\"ur
Gravitationsphysik (Albert-Einstein-Institut), Callinstr. 38,
30167 Hannover, Germany}

\author{Karsten Danzmann}
\affiliation{Institut f\"ur Atom- und Molek\"ulphysik, Universit\"at Hannover and Max-Planck-Institut f\"ur Gravitationsphysik (Albert-Einstein-Institut), Callinstr. 38, 30167 Hannover, Germany}

\author{Roman Schnabel}
\affiliation{Institut f\"ur Atom- und Molek\"ulphysik, Universit\"at Hannover and Max-Planck-Institut f\"ur Gravitationsphysik (Albert-Einstein-Institut), Callinstr. 38, 30167 Hannover, Germany}




\date{\today}

\begin{abstract}
We report on the demonstration of broadband squeezed laser beams
that show a frequency dependent orientation of the squeezing
ellipse. Carrier frequency as well as quadrature angle were stably
locked to a reference laser beam at 1064~nm. This frequency
dependent squeezing was characterized in terms of noise power
spectra and contour plots of Wigner functions. The later were
measured by quantum state tomography. Our tomograph allowed a
stable lock to a local oscillator beam for arbitrary quadrature
angles with $\pm 1^\circ \,$ precision. Frequency dependent
orientations of the squeezing ellipse are necessary for squeezed
states of light to provide a broadband sensitivity improvement in
third generation gravitational wave interferometers. We consider
the application of our system to long baseline interferometers
such as a future squeezed light upgraded GEO\,600 detector.
\end{abstract}

\maketitle

\section{Introduction}

Gravitational waves (GW) have long been predicted by Albert
Einstein using the theory of general relativity, but so far have
not been directly observed \cite{Thorne87}. Currently, an
international array of first-generation, kilometer-scale laser
interferometric gravitational-wave detectors, consisting of
GEO\,600~\cite{geo02,geo04}, LIGO~\cite{LIGO,geo04},
TAMA\,300~\cite{TAMA} and VIRGO~\cite{VIRGO04}, targeted at
gravitational-waves in the acoustic band from 10~Hz  to 10~kHz, is
going into operation. These first-generation detectors are all
Michelson interferometers with suspended mirrors. Injecting a
strong carrier light from the bright port, the anti-symmetric mode
of arm-length oscillations (e.g.\ excited by a gravitational wave)
yields a sideband modulation field in the anti-symmetric (optical)
mode which is detected at the dark output port. To yield a high
sensitivity to gravitational waves,these interferometers have long
arm lengths of 300~m up to 4~km and high circulating laser powers,
realized by additional arm cavities and power recycling
\cite{DHKHFMW83pr}. Further improvement in sensitivity can be
achieved by signal recycling \cite{Mee88}, an advanced technique
which will be implemented in second-generation GW interferometers
and is already successfully implemented in GEO\,600
\cite{Grote04,Smith04}.

\begin{figure}[ht!]
\centerline{\includegraphics[width=8.0 cm]{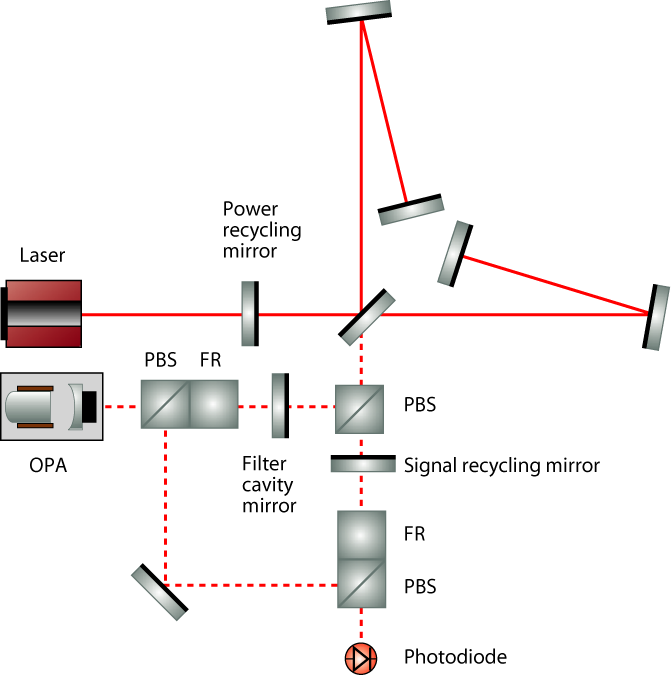}}
  \vspace{0mm}
  \caption{\coon Interferometer topology of the GEO\,600 gravitational-wave detector with
  additional filter cavity mirror, polarizing optics and squeezed light input from an optical parametric
  amplifier (OPA). Parts of the signal recycling cavity are used to form a filter
  cavity that provides the desired frequency dependence of squeezed light applied to
  the dark port as proposed in \cite{SHSD04}. For high frequencies (above 1 kHz)
  a single filter cavity is sufficient to gain a broadband sensitivity improvement
  from injected squeezed light. PBS: Polarizing Beam Splitter; FR: Faraday Rotator}
  \label{interferometer}
\end{figure}

It was first proposed by Caves~\cite{Cav81} that squeezed light
injected into the dark port of a GW interferometer can reduce the
high laser power requirements. In subsequent experiments
shot-noise reduction from squeezed light was demonstrated in
table-top interferometers \cite{XWK97,GSYL87,KSMBL02}.
Unruh~\cite{Unruh82} and others~\cite{GLe87,JRe90,PCW93,KLMTV01}
have found and proven in different ways that quantum correlated
light, e.g. squeezing at a quadrature angle of $45^\circ$, can
reduce the quantum noise below the so-called standard quantum
limit (SQL) \cite{BraginskyKhalili92} at least at some
frequencies. It was also shown that squeezed light with a
frequency dependent orientation of the squeezing ellipse can
reduce the quantum noise over the complete spectrum. In all cases
interferometer topologies without signal-recycling were
considered. Recently Harms {\it et al.} \cite{HCCFVDS03} have
generally shown that signal-recycled interferometers will benefit
from squeezed light similarly to conventional interferometers as
mentioned above. This result has strongly motivated further
investigations on squeezed light because second-generation
detectors like Advanced LIGO will combine arm cavities and
signal-recycling (then so-called resonant sideband extraction
\cite{MSNCSRWD93,HMSRWD96}) and will be quantum noise limited over
a substantial fraction of its detection spectrum. We emphasize
that injected squeezed vacuum states can lead to a broadband
sensitivity improvement, even at the two resonance frequencies of
the detuned (signal tuned) signal recycling cavity, i.e. the
optical resonance \cite{Mee88,HSMSWWSRD98,FHSMSLWSRWD00} and the
opto-mechanical (optical spring) resonance
\cite{BCh01a,BCh01b,BCh02a}. In \cite{HCCFVDS03} it was further
shown that for a fixed readout homodyne angle two detuned filter
cavities can provide the optimal frequency dependence for
initially frequency independent squeezed light. In \cite{SHSD04}
we investigated the high frequency limit of signal-recycled
gravitational wave detectors. For detection frequencies above
$\approx 1$kHz, GEO\,600 is expected to be quantum noise
(shot-noise) limited, thus a reduction of quantum noise would lead
to a sensitivity improvement without further reduction of other
noise sources, such as thermal noise. In the shot-noise limited
regime of a signal recycled interferometer a \emph{single} filter
cavity can provide the optimal frequency dependence for the
squeezing ellipse. In that case the filter cavity should exactly
match the interferometers signal recycling cavity and needs to be
locked to the same signal sideband frequency but with the opposite
sign. Fig.~\ref{interferometer} shows a proposal according to
\cite{SHSD04} of how such a filter cavity could be implemented
using parts of the signal recycling cavity and additional optics.

In this paper we present what we believe to be the first
demonstration of a squeezed source/single filter cavity system,
that provides a frequency dependent squeezed laser beam. Long term
stable control has been achieved that enabled a detailed
tomographic characterization of the nonclassical laser beam
generated. Cooperation of two identical cavities with opposite
detunings, so to speak signal recycling cavity and filter cavity,
has been inferred from our experimental data.

\section{Squeezed states of light}

A single mode of the electro-magnetic field can be described by
its field amplitude or annihilation operator $\hat{a}(\omega)$,
which has the commutation relation $[\hat a(\omega), \hat
a^\dagger(\omega)] = 1$. The non-Hermitian operator $\hat
a(\omega)$ can be decomposed according to
\begin{eqnarray}
\label{at}
     \sqrt{\frac{\omega_0+\Omega}{\omega_0}}\hat a(\omega_0+\Omega) & = & \frac{\hat a_1(\Omega) + i \hat a_2(\Omega)}{\sqrt{2}}\,, \\
     \sqrt{\frac{\omega_0-\Omega}{\omega_0}}\hat a^\dagger(\omega_0-\Omega) & = & \frac{\hat a_1(\Omega) - i \hat a_2(\Omega)}{\sqrt{2}}\,,
\end{eqnarray}
where $\hat a_{1,2}(\Omega)$ are the amplitude (sub-script 1) and
phase (sub-script 2) quadrature operators of the two-photon
formalism acting in the space of modulation frequencies $\Omega$
\cite{CSc85}. Correspondingly, the carrier frequency of the field
is denoted by $\omega_0$. The discrete commutation relation $[\hat
a_1(\Omega), \hat a^\dagger_2(\Omega)]=i$ follows directly from
the commutation relation of $\hat a(\omega)$ and $\hat
a^{\dagger}(\omega)$.  This relation places a fundamental
limitation on how well one quadrature of an optical beam can be
known, given some knowledge of the orthogonal quadrature.  This
can be expressed by the uncertainty relation
\begin{equation}
\label{uncertainty}
 \Delta \!  ^{2} \hat a_1(\Omega) \, \Delta \!  ^{2} \hat a_2(\Omega) \ge \frac{1}{4}\,,
\end{equation}
where the operator variances are denoted by
\begin{equation}
\Delta \!  ^{2} \hat a^{\phantom{\dagger}}_{1,2} =
\frac{1}{2}\langle \hat a^{\phantom{\dagger}}_{1,2}\hat
a^\dagger_{1,2}+\hat a^{\dagger}_{1,2}\hat
a^{\phantom{\dagger}}_{1,2}\rangle - |\langle \hat
a^{\phantom{\dagger}}_{1,2}\rangle|^{2}.
\end{equation}
For pure states minimum uncertainty is given by the equal sign in
Eq.~(\ref{uncertainty}).
For conciseness where possible we omit the explicit frequency
dependence $(\Omega)$ and since $\Omega\ll\omega_0$, we introduce
the following approximation to Eq.~(\ref{at})
\begin{eqnarray}
\label{at2}
     \hat a_1 & = & \frac{a+a^\dagger}{\sqrt{2}}\,, \\
     \hat a_2 & = & \frac{a-a^\dagger}{\sqrt{2i}}\,.
\end{eqnarray}
We need to keep in mind that amplitudes on the right-hand side act
at different sideband frequencies; however, for all practical
purposes, the quadrature amplitudes are Hermitian operators and
therefore will be used as representatives of measurement results.
With their continuous eigenvalue spectra these operators span the
phase space which the Wigner function is defined in. An arbitrary
quadrature in this phase space can be defined as \bea
\hat a_\theta &\!=&\frac{1}{\sqrt{2}}(\hat a e^{-i\theta} + \hat a^\dag e^{i\theta}) \;\;\\
\text{or}\;\;\;\; \hat a_\theta &\!=&\hat a_1\cos\theta+\hat
a_2\sin\theta\,. \eea

We move to the rotating frame at carrier frequency $\omega_0$
where the carrier (local oscillator) stands still on the real
axis. Using a photodiode and demodulating the photoelectric
current at $\Omega$ we actually observe the beat of the two
sidebands contrarily rotating at $\pm\Omega$ with the carrier. If
some apparatus changes the phase of only one sideband by the angle
$\Phi$ keeping the carrier as well as the other sideband
unchanged, we find that the quadrature angle of constructive
interference changes by half the amount of the single sideband
rotation \bea \label{rotation}
\hat a_{\theta^\prime}\!&\!=& \frac{1}{\sqrt{2}}(\hat a e^{i\theta} + \hat a^\dag e^{-i(\theta+\Phi)})\nonumber\\
&\!=& \frac{e^{-i\Phi/2}}{\sqrt{2}}\left(   \hat a e^{i\theta^\prime} + \hat a^\dag e^{-i\theta^\prime}   \right)\\
                              &\! &\textrm{with}\quad \theta^\prime=\theta+\Phi/2\, .\nonumber
\eea

The transfer-function $\rho$ of the reflection from a cavity with
round-trip length $2L$ and reflectivities $r_1\:\mathrm{and}\:r_2$
is given by \beq \label{rho}
\rho(\Omega)=\frac{r_1-r_2\,e^{2i(\Omega-\Omega_d)L/c}}{1-r_1r_2\,e^{2i(\Omega-\Omega_d)L/c}}\,,
\eeq where $\omega_0\,+\,\Omega_d$ is a resonance frequency. Hence
sidebands are phase shifted by the amount of $\arg(\rho)$ which
depends on sideband frequency $\Omega$. As we will see in the next
section, for our purpose we chose $r_2$ close to unity resulting
in a strongly over-coupled cavity, henceforth simply termed filter
cavity.

Now we can consider squeezing at a certain frequency generated by
the beat of two sidebands with the carrier. A squeezed state
$\ket{r_s,\,\theta,\,\alpha}$ is obtained by applying the
squeezing operator
$S(r_s,\,\theta)=\exp\left[-r_s\,(\exp(-2i\theta)\,\hat a
^2-\exp(2i\theta)\,\hat a ^\dag \hspace{0pt}^2)\right]$ to a
coherent state $\ket{\alpha}\,$ \cite{WallsMilburn}. The resulting
squeezed state comes with the desired noise power reduction of
$\exp{(-r_s)}$ in quadrature $\hat a_\theta$ compared to coherent
noise. Exactly as in classical sideband modulations the phase
relation of the sidebands and the carrier determines the
quadrature angle, here the angle of the squeezed quadrature.

In our experiment we used optical parametric amplification for
squeezed light generation. Initially the angle of the squeezed
quadrature does not depend on frequency and the variance of the
quadrature operator $\vari{\hat a_\theta}$ reads \beq \vari{\hat
a_\theta}=\cosh(2r_s)-\sinh(2r_s)\cos(2(\theta-\theta_s))\,. \eeq
The light beams noise powers are therefore white in frequency
space at all angles when leaving the source. If then the squeezed
beam is reflected from a detuned cavity having a small linewidth
compared to its detuning the squeezing angle in phasespace
$\theta_s$ will then depend on sideband frequency by virtue of
$\arg(\rho)$ in Eq.~(\ref{rho}) yielding \beq \label{rotation2}
\theta_s(\Omega)=\theta_{\mbox{\tiny{OPA}}}+\frac{1}{2}\arg(\rho(\Omega))\,.
\eeq Here $\theta_{\mbox{\tiny{OPA}}}$ is the initial, frequency
independent, but variable angle of the squeezed quadrature.

Finally we define the time dependent normalized quadrature fields
$\hat q_{1,2,\theta}(\Omega,t)$ whose variances are directly
proportional to our measurement quantities,
\begin{equation}
\hat q_{1,2,\theta}(\Omega,t) = \hat a_{1,2,\theta}(\Omega) e^{-i\Omega t} + \hat a^\dag_{1,2,\theta}(\Omega) e^{i\Omega t}\,.
\end{equation}
Note that the right-hand side of Eq.~(\ref{uncertainty}) is equal to $1$ if the quadrature fields $\hat q_{1,2}(\Omega,t)$ are considered instead of its amplitudes.

\section{Experimental}

\begin{figure}[t!]
\centerline{\includegraphics[width=8.6cm]{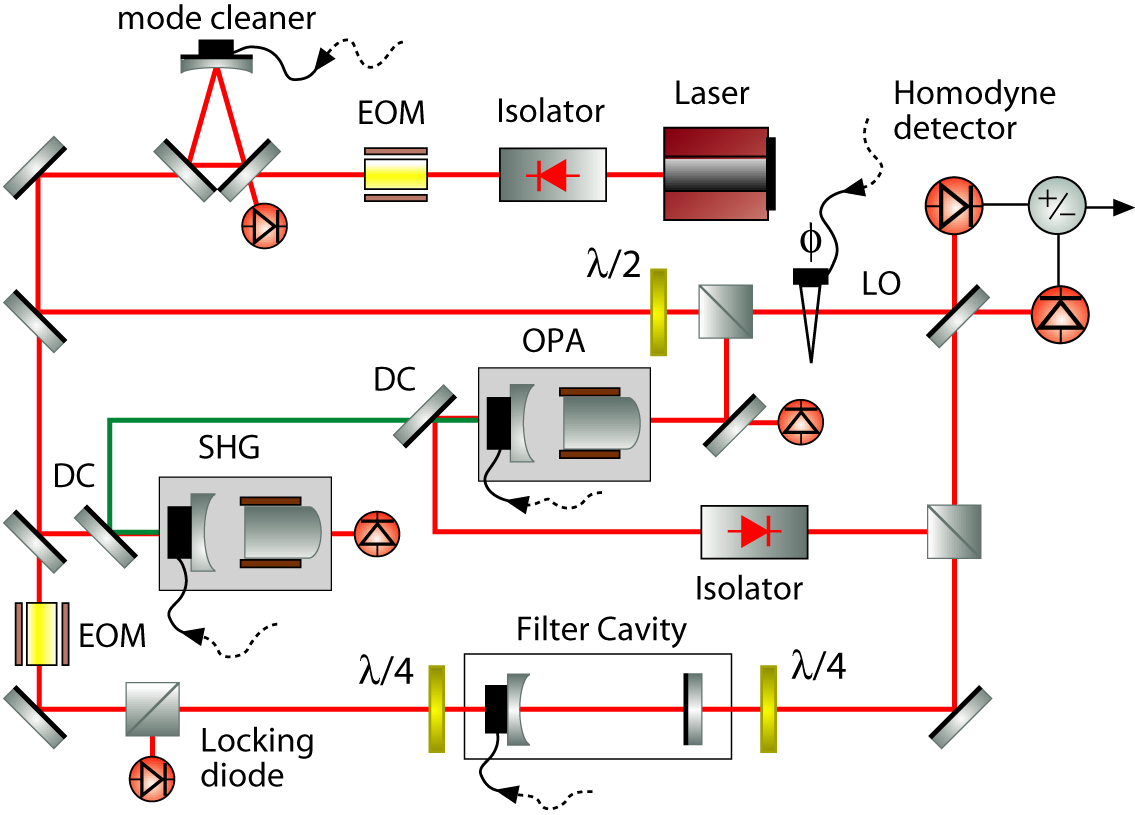}}
  \vspace{0mm}
  \caption{\coon Schematic of the experiment. Amplitude squeezed light is generated in an OPA
  cavity of controlled length. The detuned filter cavity provides frequency dependent
  squeezing suitable for a broadband quantum noise reduction of a shot-noise limited
  signal-recycled interferometer. SHG: Second Harmonic Generation;
  EOM: Electro Optical Modulator; DC: Dichroic Mirror; LO: Local Oscillator; \vrule width 3mm depth -.6mm\,: Piezo-Electric Transducer }
  \label{experiment}
\end{figure}

In our experimental setup (Fig.~\ref{experiment}) we used an
optical parametric amplifier (OPA) to produce a dim laser beam at
1064~nm carrying squeezed states at sideband frequencies from 3 up
to about $30$~MHz limited by the linewidth of the OPA cavity. The
OPA was constructed from a type I phase matched MgO:LiNbO$_{3}$
crystal inside a hemilithic (half-monolithic) resonator
\cite{SLMS98,BTSL02}. The resonator was formed by a high
reflective (HR) coated crystal surface of 10~mm radius of
curvature on one side and an externally-mounted cavity mirror of
25~mm radius of curvature and of reflectivity $r=\sqrt{0.95}$ on
the other side. The intra-cavity crystal surface was
anti-reflection coated (AR) for 1064~nm and 532~nm. The
MgO:LiNbO$_{3}$ crystal and the output coupler were separated by
an 24~mm air gap creating a cavity mode for the resonant 1064~nm
light with a 34~$\mu $m waist at the center of the crystal. The
OPA was seeded through the HR-coated surface with a coherent beam
of 15~mW at the fundamental wavelength and pumped through the
coupler with 300~mW of the second harmonic (532~nm) which results
in a classical gain of 5. The green pump light simply double
passed the nonlinear crystal because the OPA cavity was not
resonant for the second harmonic wave length. The length of the
OPA cavity was controlled using a sideband modulation technique
based on an (intra-cavity) refractive index modulation on the
MgO:LiNbO$_{3}$ crystal. This was achieved by a radio-frequency
($19.8$~MHz) electric field applied to two copper plates that were
placed on opposite sides of the nonlinear crystal. The error
signal was determined by detecting the reflected seed beam and
demodulating the resulting photocurrent at the used
radio-frequency.
From the same beam we also generated an error signal for the phase
difference of fundamental and second harmonic waves inside the
OPA. This enabled a stable lock to deamplification of the seed
beam generating a dim amplitude quadrature squeezed beam of about
70~$\mu$W at 1064~nm. This control loop used a phase modulation on
the second harmonic pump at about $19.7$~MHz that modulated the
amplification of the OPA. A stable lock to amplification is
similarly possible thereby generating a phase quadrature squeezed
beam. Note that both control locking loops did not require any
measurement on the squeezed beam leaving the OPA.
\begin{figure}[ht!]
\centerline{\includegraphics[width=8cm]{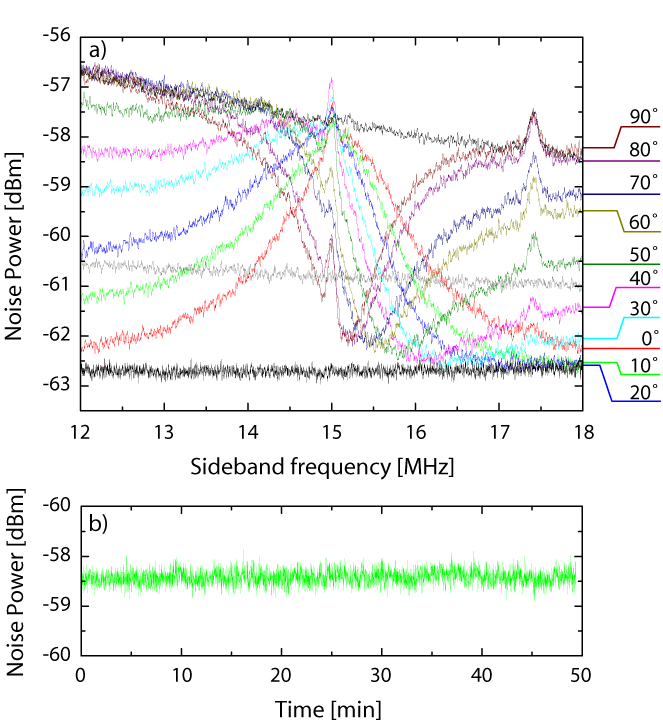}}
  \vspace{0mm}
  \caption{\coon (a) Measured noise spectra of the frequency dependent squeezed
  laser beam, stably locked to a homodyne oscillator at various quadrature
  angles. The measured quadratures show a smooth rotation from squeezed to
  anti-squeezed quantum noise or vice versa. The filter cavity was detuned
  from the carrier frequency by $\Omega_d=15.15$~MHz. (b) Demonstration of
  locking stability here for $10^\circ$ at $14.7$\,MHz.}
  \label{spectra1}
\end{figure}

\begin{figure}[ht!]
\centerline{\includegraphics[width=8cm]{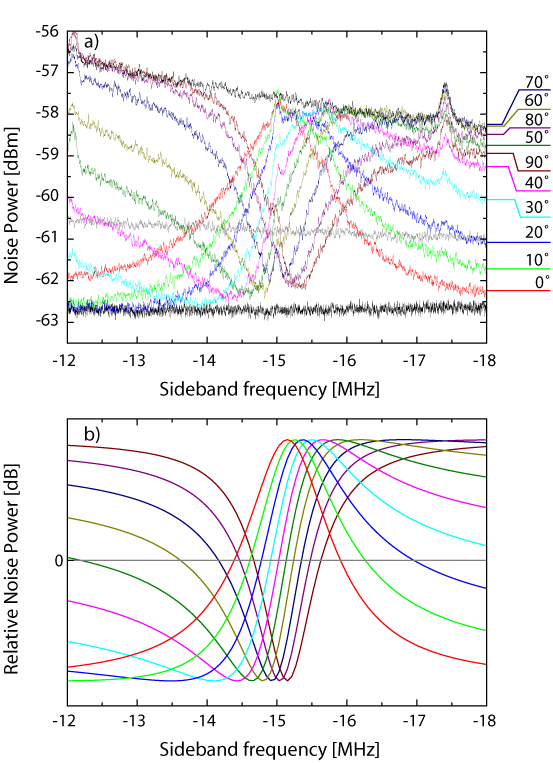}}
  \vspace{0mm}
  \caption{\coon Measured noise spectra of the frequency dependent squeezed laser beam,
  stably locked to a homodyne oscillator at various quadrature angles. The measured
  quadratures show a smooth rotation from squeezed to anti-squeezed quantum noise or
  vice versa. Here the filter cavity was detuned from the carrier frequency
  by $\Omega_d=-15.15$~MHz. (b) Comparison with theoretical curves assuming pure states and perfect mode matching into the filter cavity.}
  \label{spectra2}
\end{figure}
Common laser source of all light beams was a monolithic non-planar
Nd:YAG ring laser of 2~Watt single mode output power at 1064 nm.
Intensity noise below $2$~MHz was reduced by a servo loop acting
on the pump diode current. To reduce excess noise above $2$~MHz
the laser beam was transmitted through a mode cleaner ring cavity.
The mode cleaner output was split into two beams as shown in
Fig.~\ref{experiment}, one of these beams was mode matched into a
second harmonic generator (SHG) to produce 532~nm light which in
turn served as a pump field for our OPA. The SHG design
\cite{SBBGL00} was similar to our OPA design and provided up to
600~mW of 532~nm light with up to 65~\% conversion efficiency.
Parts of the remaining laser power were used as OPA seed
radiation, as local oscillator for homodyne detection (18~mW) and
as a filter cavity length control beam (10\,mW).

The dim squeezed laser beam from the OPA was first passed through
a Faraday isolator, preventing the OPA from any back-scattered
light. A $\lambda/4$-waveplate turned the s-polarized beam into a
circularly polarized beam which was then mode matched into our
filter cavity (FC) and retro-reflected into the homodyne detector
for quantum state tomography. According to
Fig.~\ref{interferometer} we chose a linear filter cavity design;
the cavity was composed of a plane mirror of reflectivity
$r_1=\sqrt{0.97}$ and a concave mirror of reflectivity
$r_2=\sqrt{0.9995}$; cavity length was $L=50$~cm resulting in a
linewidth of $1.47$\,MHz. The filter cavity was detuned by
$\pm15.15$~MHz throughout our measurements and stably locked to
the lower or upper sideband, respectively. Length control was
achieved by the Pound-Drever-Hall (PDH) locking technique
utilizing a circularly polarized laser beam that carried $15$~MHz
phase modulation sidebands and was coupled into the filter cavity
from the opposite side.

All spectra presented in this paper were obtained from homodyne
detector output photocurrents analyzed in a spectrum analyzer
(Rohde\&Schwarz FSP3) with 100~kHz resolution bandwidth and 100~Hz
video bandwidth over the frequency range from $12$ to $18$~MHz.
Each spectrum was at least 5~dB above the detection dark noise
which was taken into account. The homodyne detector was
constructed from two optically and electronically matched Perkin
Elmer C30619 photodiodes. The phase difference of local oscillator
and squeezed beam was controlled by using a combined DC/RF error
signal. Usually a homodyne detector is controlled for an amplitude
quadrature measurement by using RF phase modulation sidebands on
the dim (squeezed) signal beam or is controlled for a phase
quadrature measurement by utilizing the difference DC current from
the two photo diodes as the error signal. In order to characterize
quantum noise at all sorts of quadrature angles we required the
optical phase at the homodyne beamsplitter to be controllable at
arbitrary phase angles. This was achieved by a sophisticated
electronic control loop. As usual the actuator was a
piezo-electric transducer shifting the optical pathlength of the
local oscillator and the locking loop was closed by a PI element.
The
 appropriate error signal (S$_{\textrm{\small{err}}}$) was gained by a linear combination of the two linearly
 independent signals from above (RF$(\phi)$ and DC$(\phi)$) which both sinusoidally
 depend on the phase angle $\phi$.
\beq \rm S_{err}(\phi)=b\cdot \rm DC(\phi)\,+\,(1-b)\cdot \rm RF
(\phi). \label{S_err}\eeq The parameter $b$ ranging from zero to
one was computer controlled. The mapping of $\phi$ on $b$ is
bijective in intervals of $[n\frac{\pi}{2},(n+1)\frac{\pi}{2})$.
Those intervals could be appended by inverting the underlying
signals to cover the whole range from 0 to $2\pi$ thereby
providing appropriate error signals for arbitrary quadrature
angles.

In Fig.~\ref{spectra1}a) and \ref{spectra2}a) we have plotted
spectral noise powers of our frequency dependent squeezed laser
beam for various quadrature angles (see legends).
Figure~\ref{spectra1}a) shows spectra for a $+15.15$~MHz detuned
filter cavity; \ref{spectra2}a) shows spectra when the cavity was
detuned to $-15.15$~MHz. The grey horizontal line represents the
shot-noise limit in our measurement and was measured by blocking
the squeezed beam. In both figures the array is bounded by
measured traces of frequency independent squeezing (lower black)
or anti-squeezing (top black). It can be seen that at high
frequencies squeezing degraded due to the bandwidth of the OPA
cavity. The frequency independent squeezed/anti-squeezed states
were measured when a beam dump was placed inside the filter
cavity. Thereby  3$\%$ loss on the squeezing due to the filter
cavity input coupler is introduced. The effect of another 3$\%$
loss can not be seen in the spectra. Note that all spectra in
Fig.~\ref{spectra1}a) and \ref{spectra2}a) were recorded by a
conventional homodyne detector on a fully locked experiment.
Figure~\ref{spectra1}b) presents the locking stability. Here we
locked the homodyne detector to measure the quadrature at about
$10^\circ$ and recorded a zero span measurement at $14.7$~MHz that
last over 50 minutes.

\begin{figure}[t!]
\centerline{\includegraphics[width=8cm]{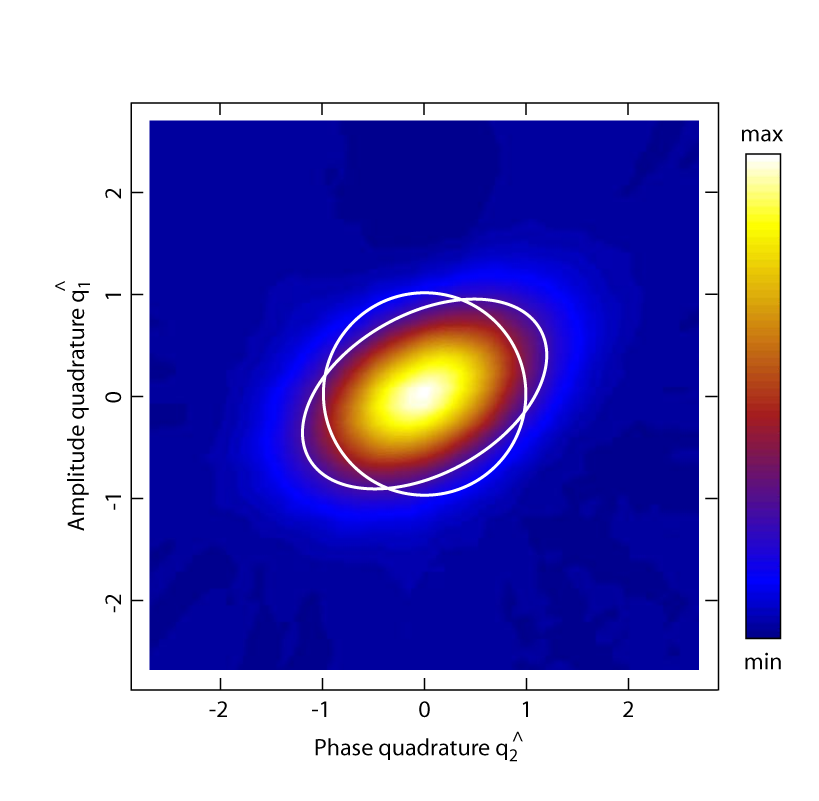}}
  \vspace{0mm}
  \caption{\coon Measured contour plot of the Wigner function at $14.1$~MHz
  sideband frequency revealing the so-called squeezing-ellipse.
  The white ellipse represents the standard deviation of the quantum
  noise; the white circle represents the size of the reference vacuum state.
  It is clearly visible that this state shows quantum correlations between
  phase and amplitude quadratures, i.e. squeezing at an angle of, here, about $40^\circ$.}
  \label{contourplot}
\end{figure}

\begin{figure}[ht!]
\centerline{\includegraphics[width=8.5cm]{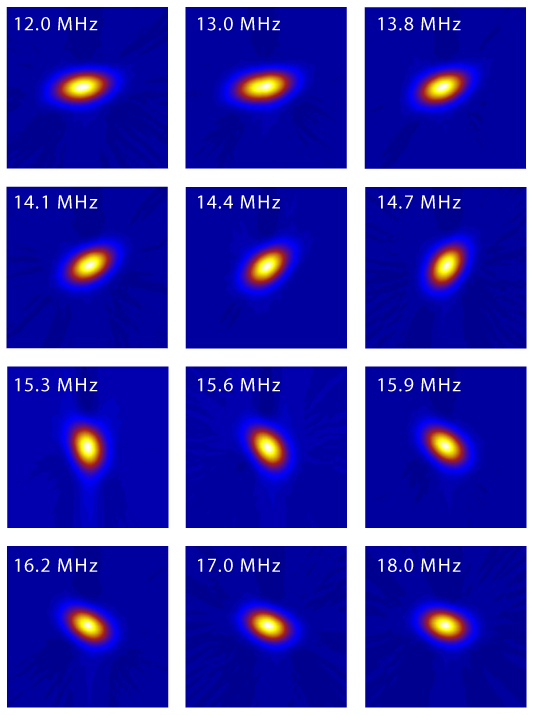}}
  \vspace{0mm}
\caption{\coon Contour plots of the squeezing ellipse at various
sideband frequencies around $15$~MHz, measured on the same
continuous wave laser beam.
  At $12$~MHz the squeezed quadrature was slightly off the vertical
  direction (amplitude quadrature); at higher frequencies the
squeezing slowly rotates into the phase quadrature at $15$~MHz
(not shown). For sideband frequencies far above $15$~MHz the
squeezing will be measured again in amplitude quadrature.
   For each tomographic measurement the laser beam was phase locked to a reference beam and the quadrature
  angle stably controlled and stepwise rotated. The phase reference was given by a
  phase modulation at $19.8$~MHz.}
  \label{contourplots}
\end{figure}

In an alternative experiment the frequency dependent squeezed
laser beam was detected by a homodyne quantum tomograph. Again the
homodyne detector described above was used to turn the laser beams
electric field properties at some quadrature angle into an
electric current. Contrary to above, now the current was not fed
into a spectrum analyzer but was mixed down with an electronic
local oscillator at some radio-frequency $\Omega$ and low-pass
filtered. The final electric signal corresponds to a time series
of measurement results on the chosen quadrature of the quantum
state at sideband frequency $\Omega$. Quantum state tomography  is
a method to reconstruct an image starting from a set of its
projections with rotating axis via a back transform algorithm. In
our case the object to reconstruct is the Wigner function of the
underlying quantum state and the projections are the probability
densities at different quadrature angles. The transformation used
is the inverse Radon-Transform which is readily implemented in
\textsc{MatLab}. As described above our setup enabled a stably
controlled measurement on arbitrary quadrature angles with a
precision of $\pm 1^\circ$. This provides a solid basis for
quantum state tomography. A \textsc{LabView} program runs the
following procedure for a predetermined set of angles:\\[-5mm]
\begin{enumerate}
\item Set $b$ parameter (see Eq. (\ref{S_err})) for the next quadrature angle to be measured.\\[-6mm]
\item Wait half a second for the control loop to stabilize.\\[-6mm]
\item Acquire measurement values on the locked quadrature.\\[-6mm]
\item Build histogram from data.\\[-5mm]
\end{enumerate}
The Wigner function was calculated afterwards from the set of
histograms. In Fig.~\ref{contourplot} the result is shown for the
quantum state at $14.1$~MHz sideband frequency (for discussion see
next section). The Resolution bandwidth was set to 100~kHz. For
reconstruction altogether 100000 quadrature values were measured
divided up on 100 equidistant quadrature angles. To characterize
our frequency dependent laser beam we measured 24 Wigner functions
at various sideband frequencies. For these measurements the
detuned filter cavity was locked to sideband frequencies of
$\Omega_d=\pm15.15$~MHz. The results for the lower sideband
($\Omega_d=-15.15$~MHz) are shown in Fig.~\ref{contourplots}.

\section{Results and Discussion}

The colored spectra in Fig.~\ref{spectra1}a) and \ref{spectra2}a)
clearly show that the squeezed laser beam investigated carried
frequency dependent squeezing. Homodyne detection of various
quadrature angles revealed that squeezed quadratures appeared at
different sideband frequencies in a spectrum of $12$ to $18$~MHz.
For comparison the lowest and highest traces (in black) show
common broadband squeezing and anti-squeezing, respectively, where
the squeezing ellipse does not rotate along the spectra of quantum
states. In Fig.~\ref{spectra2}a) and \ref{spectra2}b) we compare
our results with theoretical curves assuming pure states, no
optical losses and perfect mode matching. Both pictures show a
good qualitative agreement. Quantitative differences can be
explained by overall losses on the squeezed beam, additional phase
noise and non perfect mode matching of the squeezed beam into the
filter cavity of  $94\%$. Especially those quadratures that show
squeezing around the detuning frequency of $15.15$~MHz are
strongly effected by non perfect mode matching. The unmatched
fraction is then anti-squeezed and significantly deteriorates the
achievable quantum noise reduction. The phase modulation spikes at
about $12.2$~MHz and $17.4$~MHz were strayfields from other
experiments and picked up by the OPA. Due to thermal drifts in our
polarizing optics the $15$~MHz modulation for filter cavity
control also leaked into our homodyne detector. Although the
signal could be completely cancelled on short time scales by fine
adjustment of the polarizing optics, the modulation frequency of
$15$~MHz used for locking the filter cavity deteriorated our
measurement which last $30$~min for a complete set of spectra. To
keep the minima in Fig.~\ref{spectra1}a) and ~\ref{spectra2}a)
clearly visible and not hidden from the $15$~MHz modulation peak,
we used an electric offset on the PDH error signal to lock the
filter cavity at a sideband frequency of $15.15$~MHz instead of
$15$~MHz.

The overall squeezing in our measurement was detected to be 2 dB
below shot-noise. This value was primarily determined by optical
losses. Compared to our previous squeezing results
\cite{SVFCGBBLD04} in this experiment the losses were even higher
due to the additional Faraday rotator and the filter cavity. Loss
contributions from photodiode efficiencies, homodyne detector
visibility, propagation losses and OPA escape efficiency gave an
overall loss on the squeezed states of 42~\%.

\begin{figure}[ht!]
  \centerline{\includegraphics[width=8cm]{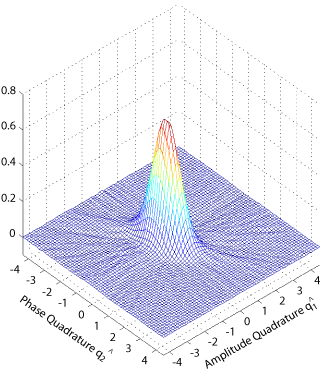}}
  \vspace{0mm}
  \caption{\coon Example of three-dimensional Wigner function plot at $12.0$~MHz
  sideband frequency according to measurement data in Fig.~\ref{contourplots}.
  Comparing states at different frequencies, frequency dependent squeezing is
  indicated by rotation along the vertical axis.}
  \label{Wignerfunctions}
\end{figure}

\begin{figure}[ht!]
\centerline{\includegraphics[width=8cm]{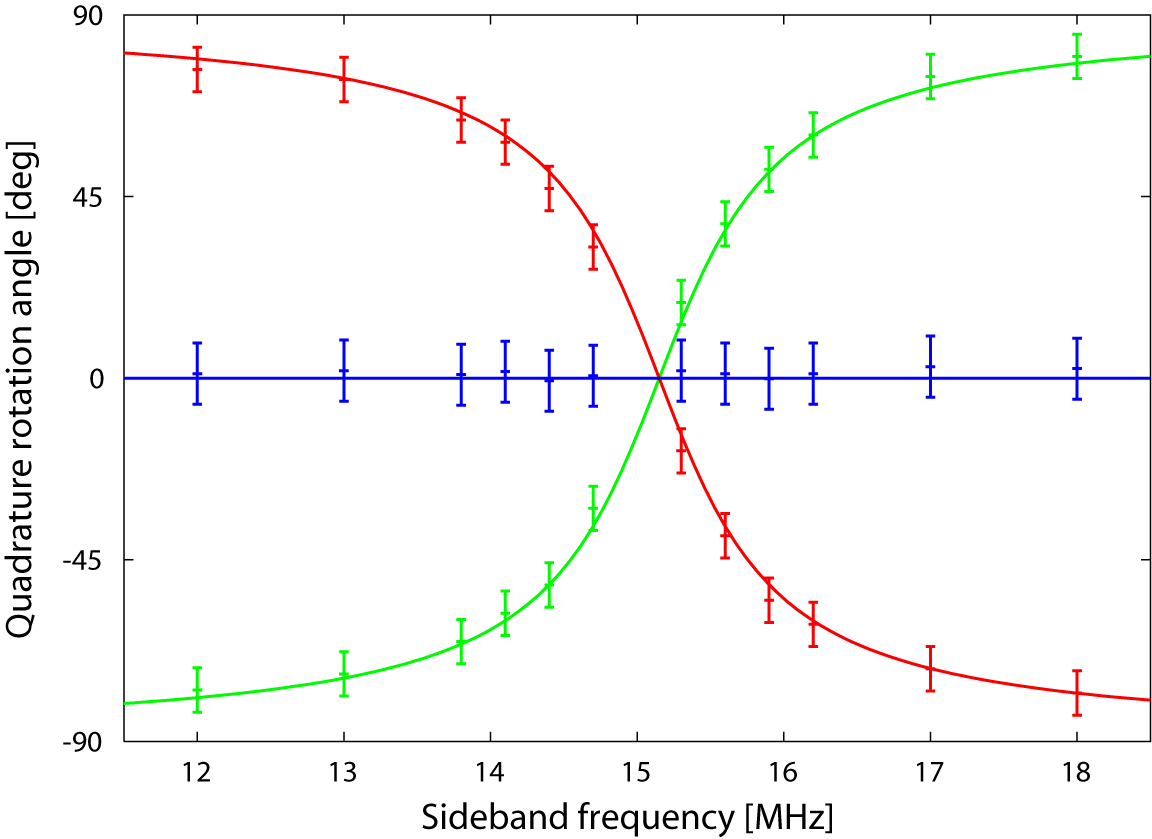}}
  \vspace{0mm}
  \caption{\coon Measured rotations of squeezing ellipses. The two curves
  show the effects from $+15.15$~MHz (green) and from $-15.15$~MHz (red) detuned filter
  cavities. One of the cavities can be regarded as a signal recycling
  cavity of a shot-noise limited interferometer.
  After two reflections the squeezing is again observed in the initial
  frequency independent quadrature. The broadband squeezed quantum noise
  can then be adjusted to the signal carrying quadrature giving a broadband
  sensitivity improvement for conventional homodyne read out.}
  \label{phase}
\end{figure}

An even clearer demonstration of frequency dependent squeezing
carried by our laser beam can be given in terms of Wigner function
contour plots. In addition to the spectra discussed above, we did
tomographic quantum state measurements \cite{Leonhardt97,BSc97} at
frequencies between 12 and $18$~MHz as described in the previous
section. Figure~\ref{contourplot} presents a measured contour plot
of squeezing-ellipse at a sideband frequency of $14.1$~MHz. The
white ellipse represents the standard deviation of the quantum
noise; the white circle represents the size of the reference
vacuum state, normalized to unity. It is clearly visible that this
state shows quantum correlations between phase and amplitude
quadratures, i.e. squeezing at about $40^\circ$. Note that here
the Wigner function is centered around the origin saying that no
coherent excitation of the modulation sideband was present.
However, additional noise in the anti-squeezed quadrature was
clearly detected, since the area of the ellipse was determined to
be 1.16 which can partly be attributed to optical losses on the
squeezing and additional phase noise. Figure~\ref{contourplots}
shows our tomography results at various sideband frequencies using
a detuned filter cavity with $\Omega_d=-15.15$~MHz. For each
tomographic measurement the homodyne detector was subsequently
locked to 100 equidistant quadrature angles for a certain
measurement time interval. An equivalent sequence was measured
using a $+15.15$~MHz detuned filter cavity. The results are not
explicitly shown but were used to plot the green curve in
Fig.~\ref{phase}. To the best of our knowledge we report on the
first fully locked tomographic characterization of nonclassical
states. In previous works the quadrature angle were slowly scanned
thereby averaging over a finite range of angles. In
Fig.~\ref{Wignerfunctions} an example of a three-dimensional
Wigner function plot is given showing the state at $12.0$~MHz
sideband frequency according to measurement data in
Fig.~\ref{contourplots}.
Fig.~\ref{phase} shows the effect of reflections from single detuned
filter cavitities on the quadrature angle. Displayed are measured
rotations of the squeezing ellipse deduced from our contour plots
for $+15.15$~MHz and $-15.15$~MHz detuning. Altogether 24 measured
Wigner function contour plots have been evaluated. For every plot
we determined the rotation angle of the squeezed quadrature with
reference given by the initially squeezed amplitude quadrature.
The two curved solid lines represent the theory according to
Eq.~(\ref{rotation2}). The very good agreement between
experimental data and theory suggests that the approximation for
small filter cavity linewidth in Eq.~(\ref{rotation2}) is valid
for our experiment. From theory we expect that in any case the sum
of both curved traces cancel to zero shown as a horizontal, blue line in
Fig.~\ref{phase} . This is indeed shown by our data points,
thereby experimentally proving that two subsequent filter cavities
that are symmetrically detuned by in our case $\pm15.15$~MHz from
the carrier, lead to then an overall identical phase shift for
carrier light and for all sideband frequencies. The
effects of both cavities cancel each other.

Figure~\ref{phase} is motivated by the fact that detuned cavities
have become widely accepted in future gravitational wave
interferometers. As said in the introduction the GEO\,600 detector
already uses a carrier detuned (signal tuned) so-called signal
recycling cavity. The second generation detector Advanced LIGO
will use a detuned signal extraction cavity. It is very likely
that also third generation detectors will make use of detuned
cavities. Our research aimed for third generation GW
interferometers where the ordinary coherent vacuum entering the
interferometers dark port is replaced by a squeezed vacuum.
Injected squeezed states will be reflected from the detuned cavity
of the interferometer and thereby will gain a frequency dependent
phase shift. As a result the squeezed quadrature will not beat
with the signal at all frequencies resulting in a reduced detector
bandwidth. Generically this can be circumvented by two or more
appropriate filter cavities which have been first proposed by
\cite{KLMTV01} to enable sensitivities below the interferometers
standard quantum limit. Contrary to \cite{KLMTV01} where ring
cavities were proposed we used a linear filter cavity in
correspondence to the design in Fig.~\ref{interferometer}
targeting the interferometers shot-noise limit. For frequencies
above 1 kHz the GEO\,600 detector is not significantly effected by
radiation pressure noise and a broadband sensitivity improvement
from squeezed light can be achieved by a single detuned filter
cavity. If the squeezed vacuum is reflected from a filter cavity
with inverted detuning compared with the interferometers recycling
cavity, both effects cancel. Then the squeezed states that are
reflected from the interferometer show a frequency independent
orientation regarding the signal field and the shot-noise limited
sensitivity of the interferometer is improved over  the complete
spectrum according the degree of squeezing achieved.

\section{Conclusion}

Our results demonstrate frequency dependent squeezing at sideband
frequencies between $12$ and $18$~MHz. We have demonstrated a
stably controlled system in a bench top experiment. Broadband but
frequency independent squeezing was provided by an optical
parametric amplifier, frequency dependent orientation of the
squeezing ellipse was achieved by reflecting the squeezing ellipse
at a detuned filter cavity. For shot-noise limited signal recycled
interferometers with homodyne readout at a fixed angle our scheme
provides a broadband sensitivity improvement. Required cooperation
of two identical cavities with opposite detunings, so to speak
signal recycling cavity and filter cavity, has been inferred from
our experimental data. We used carrier light at 1064~nm which is
the wavelength for current gravitational wave interferometers;
squeezing at acoustic frequencies was recently demonstrated for
the same carrier wavelength. \cite{MGBWGML04}. Our scheme can be
scaled to long baselines and acoustic
frequencies using current technology.\\

This work has been supported by the Deutsche
Forschungsgemeinschaft and is part of Sonderforschungsbereich 407.
We also acknowledge Jan Harms for his contributions to the theory
section and Joshua R. Smith for his invaluable suggestions that
contributed to the form and clarity of this paper.

\appendix

\end{document}